\documentclass[doublecol]{epl2}

\title{Many-body quantum coherence and interaction blockade in Josephson-linked Bose-Einstein condensates}
\shorttitle{Many-body quantum coherence and interaction blockade}

\author{Chaohong Lee\inst{1}\thanks{Corresponding author's email:\email{chl124@rsphysse.anu.edu.au}}
\and Li-Bin Fu\inst{1,2} \and Yuri S. Kivshar\inst{1}}
\shortauthor{C. Lee \etal}

\institute{\inst{1}Nonlinear Physics Center and ARC Center of
Excellence for Quantum-Atom Optics, Research School of Physical
Sciences and Engineering, Australian National University, Canberra
ACT 0200, Australia\\
\inst{2} Institute of Applied Physics and Computational Mathematics,
P.O. Box 100088, Beijing, P. R. China}


\abstract{We study many-body quantum coherence and interaction
blockade in two Josephson-linked Bose-Einstein condensates. We
introduce universal operators for characterizing many-body
coherence without limitations on the system symmetry and total
particle number $N$. We reproduce the results for both coherence
fluctuations and number squeezing in {\em symmetric} systems of
large $N$, and reveal several peculiar phenomena that may occur in
{\em asymmetric} systems and systems of small $N$. For asymmetric
systems, we show that, due to an interplay between asymmetry and
inter-particle interaction, the coherence fluctuations are
suppressed dramatically when $|E_{C}/E_{J}|\ll 1$, and both {\it
resonant tunneling} and {\it interaction blockade} take place for
large values of $|E_{C}/E_{J}|$, where $E_C$ and $E_J$ are the
interaction and tunneling energies, respectively. We emphasize
that the resonant tunneling and interaction blockade may allow
creating single-atom devices with promising technology
applications. We demonstrate that for the systems at finite
temperatures the formation of self-trapped states causes an
anomalous behavior.}

\begin{document}

\maketitle

Utilizing the current experimental techniques, it is possible to
prepare atomic Bose-Einstein condensates in which all atoms occupy
the same single-particle state. To explore the quantum coherence
between two such condensates, it is natural to link them through
the Josephson tunneling~\cite{RMP-Leggett}, and then release for
producing interference. Atomic
interferometry~\cite{Science-atomic-coherence-Kasevich} provides
an important experimental tool for the fundamental studies in
atomic, optical, and quantum physics, as well as promising
applications in high-precision measurements and quantum
information processing \cite{Nature-Phillips}. By loading
condensates into double-well potentials, atomic interferometers
have been realized experimentally
~\cite{DWI-MIT,DWI-Heidelberg,DWI-Austria,DWI-SUT}. Furthermore,
some many-body quantum effects such as coherence
fluctuations~\cite{DWI-Austria,ThermalDWI-Heidelberg}, conditional
tunneling~\cite{ICOLS2007-Bloch}, number
squeezing~\cite{Science-squeezing-Kasevich}, and long-time
coherence assisted by number squeezing~\cite{LTC-MIT} as well as
finite-temperature effects~\cite{ThermalDWI-Heidelberg} have been
observed in experiment. However, clear theoretical explanation for
many of those effects, such as the full physical picture of the
coherence fluctuations~\cite{DWI-Austria,ThermalDWI-Heidelberg},
the resonant tunneling and the interaction blockade
\cite{Coulomb-blockade,Coulomb-blockade-book,Coulomb-blockade-application,Coulomb-blockade-application-book}
are still largely missing. Below, we present a general approach
for analyzing many-body quantum coherence and interaction blockade
and give a full picture of the coherence fluctuations.
Particularly, our results of resonant tunnelling can qualitatively
explain the conditional tunneling which has been observed recently
~\cite{ICOLS2007-Bloch}.

To study quantum fluctuations in the Josephson-linked coherent
systems such as Bose-Einstein condensates in double-well
potentials, one has to analyze the system within the full quantum
theory. Under the conditions of tight binding, the quantum system
can be mapped onto a two-site Bose-Hubbard
model~\cite{Science-squeezing-Kasevich}. Then, the number
fluctuations can be easily analyzed by applying the techniques
well-developed in quantum optics~\cite{Nature-squeezing-Walls}.
Where, however, the coherence fluctuations could not be clearly
depicted. Previously, the phase fluctuations have been used as a
measure characterizing the coherence fluctuations. However, due to
the absence of the phase operators in the Bose-Hubbard
Hamiltonians, one has to introduce the quantum-phase
models~\cite{Q-phase-models,GQ-phase-models}. The quantum-phase
models describe qualitatively the systems with small number
differences, but they can not describe the systems of large number
differences and fail to maintain the hermiticity for asymmetric
systems. Additionally, because the phase is measured modulo
$2\pi$, the fluctuations of phase do not provide the best
quantitative characterization for the coherence
fluctuations~\cite{LTC-MIT}.

In this Letter, we introduce Hermitian operators for the phase
coherence rather than a relative phase, and then analyze the
coherence fluctuations and interaction blockade. The coherence
operators are defined with the atomic creation, annihilation, and
number operators, so that they can be calculated directly with the
general Bose-Hubbard models. The operational approach has two
advantages: first, its validity is independent of the total
particle numbers and, second, it works for both symmetric and
asymmetric systems. We examine how the coherence fluctuations
depend on the total number of particles, the inter-particle
interaction energy, the junction tunneling energy, asymmetry, and
temperature. Without loss of generality, for all calculations
about the fluctuations in zero-temperature systems, we only
consider the ground states of the full quantum model: quantized
Bose-Josephson junction. For the calculations about the
fluctuations in finite-temperature systems, we use both the full
quantum and the mean-field models.

Below, we not only reproduce the pictures of coherence
fluctuations and number squeezing in large-size symmetric systems
which can be described by the quantum phase models, but also
reveal several peculiar phenomena in small-size systems and
asymmetric systems which can not be described by the standard
quantum phase models. Such as, (i) due to an interplay between the
system asymmetry and inter-particle interaction, single-atom
resonance effects suppress the quantum fluctuations of coherence
and the interaction blockade occurs between the neighboring
resonant peaks; and (ii) the occupation of highly-excited states
of degeneracy (macroscopic quantum self-trapping) causes an
anomalous behavior of the coherence fluctuations at finite
temperatures. These results are qualitatively consistent with the
recent experiment data on many-body quantum
phenomena~\cite{DWI-Austria,ThermalDWI-Heidelberg,Science-squeezing-Kasevich,LTC-MIT}
and finite-temperature dynamics~\cite{ThermalDWI-Heidelberg}. In
particular, the resonant effects are confirmed by the most recent
experimental observations of the conditional tunneling
\cite{ICOLS2007-Bloch}; the physics of the interaction blockade
and resonant tunneling open a route for designing single-atom
devices for applications in sensitive metrology and information
technology.

We consider the quantum model of two Josephson-linked atomic
condensates with the conserved total number of particles. From the
physical point of view, this situation can be realized by confining
atomic condensates in a double-well
potential~\cite{DWI-MIT,DWI-Heidelberg,DWI-Austria,DWI-SUT}.
Eliminating the constant terms including the total number of atoms,
the system Hamiltonian
\begin{equation}
H = - J (a_{1}^{+}a_{2} + a_{1}a_{2}^{+}) + \frac{\delta}{2}(n_2 -
n_1) + \frac{E_C}{8}(n_2 - n_1)^2
\end{equation}
describes a quantized Bose-Josephson junction. Here, $a_{j}^{+}
(a_{j})$ are the atomic creation (annihilation) operators for the
$j$-th ($j=1,2$) well, and $n_j$ are the corresponding number
operators. Thus, the system behavior is determined by the total
number $N$, the charging (inter-particle interaction) energy $E_C$,
the inter-well tunneling strength $J$, and the asymmetry parameter
$\delta$. It has been suggested that the symmetric systems with
negative charging energies can be used to realize the
Heisenberg-limited interferometry~\cite{MZ-interferometry-Lee}.
Here, we consider both symmetric and asymmetric systems with the
positive charging energies.

Under the mean-field approximation, $a_j \to \sqrt{n_j} \,\,
\textrm{e}^{i\phi_j}$ and $a_j^+ \to \sqrt{n_j} \,\,
\textrm{e}^{-i\phi_j}$, the macroscopic quantum behavior (such as
Josephson oscillations, self-trapping~\cite{MQST-Smerzi}, and
dynamical bifurcation~\cite{bifurcation-Lee}) obeys a classical
non-rigid pendulum Hamiltonian,
\begin{equation}
H_{cl} = E_{C} n^{2} / 2 + \delta n - E_{J} \sqrt{1-\left ( 2n/N
\right )^{2}} \cos \phi, \label{eq_H}
\end{equation}
with the quasi-angular momentum $n = (n_2-n_1)/2$ and the relative
phase $\phi = \phi_2 - \phi_1$. Here, $E_J = NJ$ denotes the
junction energy. For the symmetric systems ($\delta = 0$) with small
number differences ($\left|2n/N\right| \sim 0$), one can introduce
the quantum phase model via $n \to i\frac{\partial}{\partial \phi}$
and $(2n/N) \to 0$ (see Ref.~\cite{Q-phase-models,GQ-phase-models}
and references therein). However, it can not depict the behavior of
systems with large number differences or self-trapping, and it fails
to maintain hermiticity when $\delta \ne 0$.

To explore the coherence fluctuations, we introduce operators for
the phase coherence. Drawing upon the quantum phase
concept~\cite{q-phase} for single-mode coherence, we define the
operators for two-mode first-order coherence as
\begin{eqnarray}
\cos \phi &=& \frac{(a_{2}^{+}a_{1} +
a_{2}a_{1}^{+})}{\sqrt{2\left\langle 2n_{1}n_{2}+n_{1} + n_{2}
\right \rangle}},
\nonumber \\
\sin \phi &=& \frac{i(a_{2}^{+}a_{1} -
a_{2}a_{1}^{+})}{\sqrt{2\left\langle 2n_{1}n_{2}+n_{1} + n_{2}
\right \rangle}} .
\end{eqnarray}
These two operators satisfy the conditions $\left \langle
\sin^{2}\phi + \cos^{2}\phi \right\rangle=1$. Without considerations
in total particle number conservation and inter-particle
interaction, similar operators have been used to study the phase
coherence of photons from different
lasers~\cite{Mandel_QuantumPhase}. So that the previous results of
photons can not be applied to systems of conserved total particle
numbers and inter-particle interaction, such as the Josephson-linked
condensates considered in this paper. Similarly, one can define the
operators for two-mode second-order coherence, $\cos (2\phi) =
\left[(a_{2}^{+}a_{1})^{2} + (a_{2}a_{1}^{+})^{2} \right]/K_S$ and
$\sin (2\phi) = i \left[(a_{2}^{+}a_{1})^{2} -
(a_{2}a_{1}^{+})^{2}\right]/K_S$, where the constant $K_S =
\left\langle 2n_{1}n_{2}+n_{1}+n_{2}\right\rangle$. Fluctuations of
the two-mode first-order coherence are quantitatively characterized
by the variance
\begin{equation}
\Delta (\cos \phi) = \left\langle \cos ^{2}\phi\right\rangle
-\left\langle \cos \phi\right\rangle ^{2}.
\end{equation}
Here, the expectation for $\cos \phi$ is
\begin{equation}
\left\langle \cos \phi \right\rangle =\frac{\left\langle
a_{2}^{+}a_{1}+a_{2}a_{1}^{+}\right\rangle }{\sqrt{2\left\langle
2n_{1}n_{2}+n_{1}+n_{2}\right\rangle }},
\end{equation}
and the expectation for $\cos^{2} \phi$ is
\begin{equation}
\left\langle \cos ^{2}\phi \right\rangle =\frac{1}{2} +
\frac{\left\langle (a_{2}^{+}a_{1})^{2}\right\rangle +\left\langle
(a_{2}a_{1}^{+})^{2}\right\rangle }{2\left\langle
2n_{1}n_{2}+n_{1}+n_{2}\right\rangle }.
\end{equation}
The variance $\Delta (\cos \phi)$ vanishes when two condensates
becomes phase-locked. Otherwise, the variance approaches the value
$1/2$ when two condensates have no phase correlations. The above
variance has been successfully employed for explaining the q-vortex
melting~\cite{q-vortex-Lee}.

\begin{figure}[ht]
\rotatebox{0}{\resizebox *{\columnwidth}{6.8cm} {\includegraphics
{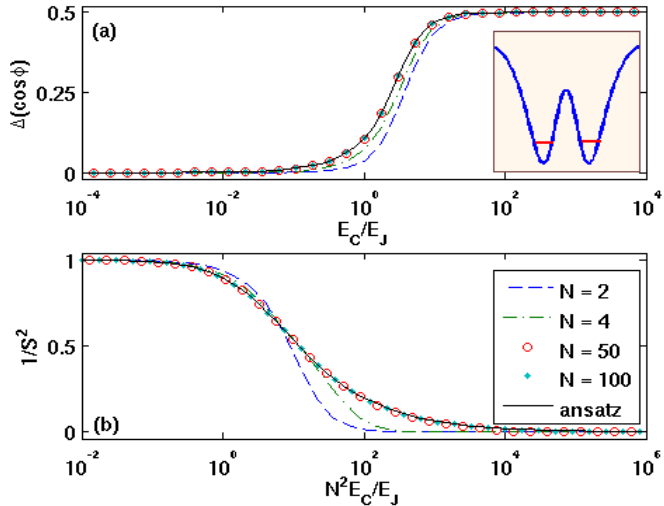}}} \caption{Coherence
fluctuations and number squeezing in symmetric Bose-Josephson
junctions. (a) Coherence fluctuations $\Delta(\cos \phi)$ vs.
$E_C/E_J$; and (b) inverse square of number squeezing $1/S^{2}$
vs. $N^2 E_C/E_J$. Dashed lines, dashed-dot lines, circles, and
dots correspond to the cases $N = 2, 4, 50$ and $100$,
respectively. For sufficiently large $N$, the curves approaches
the result obtained from the Gaussian ansatz applied to the
probability amplitudes (solid line).} \label{fig1}
\end{figure}

For symmetric systems at zero temperature, the coherence
fluctuations $\Delta (\cos \phi)$ and the number squeezing $S =
\sqrt{N/\Delta N_r}$ (where $\Delta N_r = \langle N_r^2 \rangle -
\langle N_r \rangle^2$ is the variance for relative numbers $N_r =
n_2 - n_1$) show different parametric dependencies. The coherence
fluctuations grow monotonously with the ratio $E_C/E_J = E_C
/(NJ)$, however, the number squeezing grows monotonously with the
ratio $N^2E_C/E_J = NE_C/J$. The behavior of the coherence
fluctuations is qualitatively consistent with the experimental
observations made via adjusting the separation
distance~\cite{DWI-Austria} or the barrier
height~\cite{ThermalDWI-Heidelberg}, which corresponds to varying
$E_J$. Because the coherence time grows with the number
squeezing~\cite{Gaussian-ansatz-You}, the different parametric
dependence of the coherence fluctuations and number squeezing
indicates that one can increase the coherence time without loss of
coherence, by increasing the total number $N$ or with some loss of
coherence, by increasing the ratio $E_C/E_J$. This result
justifies qualitatively the experimental approach for adjusting
the coherence time via varying the barrier height (i.e.,
controlling $E_J$)~\cite{LTC-MIT}, in which the coherence time
grows with the barrier height. For sufficiently large $N$, the
parametric dependence of coherence fluctuations and number
squeezing are consistent with the previous results obtained from
other
approaches~\cite{Q-phase-models,Gaussian-ansatz-You,BEC-book}. Our
operational approach also explores the finite-size effects in
systems of small $N$. For commensurate systems (whose $N$ are even
integers) of very weak or strong interactions, the coherence
fluctuations are almost independent of $N$. However, in the
intermediate region of $E_C/E_J \sim 1$, the coherence
fluctuations decrease with $N$. In Fig.~\ref{fig1}, we show the
parametric dependence for the ground-state coherence fluctuations
and number squeezing for symmetric systems with different total
number of particles.

\begin{figure}[ht]
\rotatebox{0}{\resizebox *{\columnwidth}{6.8cm} {\includegraphics
{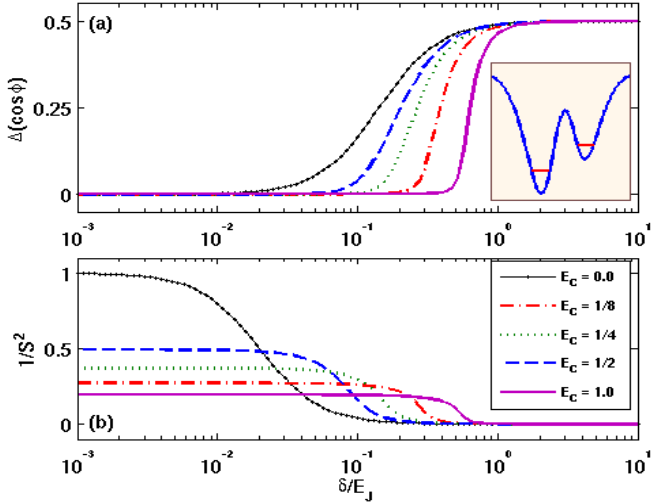}}} \caption{Coherence fluctuations and
number squeezing in asymmetric systems with $N = 100$, $E_J =
100$, and $E_C/E_J \ll 1$. (a) Coherence fluctuations vs.
$\delta/E_J$; and (b) inverse square of number squeezing $1/S^{2}$
vs. $\delta/E_J$.} \label{fig2}
\end{figure}

For asymmetric systems at zero temperature and without
inter-particle interaction ($E_C=0$), both coherence fluctuations
and number squeezing  increase monotonically with the asymmetry
parameter $\delta$. For small values of $E_C/E_J$, before the
asymmetry $\delta$ starts dominating the dynamics, the coherence
fluctuations can be suppressed by an interplay between asymmetry and
repulsive inter-particle interaction; the suppression becomes more
significant for larger values of $E_C$. This suppression of the
coherence fluctuations originates from the different localization
mechanisms dominated by asymmetry and repulsive inter-particle
interaction. The asymmetry tends to localize all atoms in the lower
well, however, the repulsive inter-particle interaction (positive
charging energy) tends to localize atoms in two wells without number
difference. In Fig.~\ref{fig2}, we show the coherence fluctuations
and number squeezing in asymmetric systems with $N = 100$, $E_J =
100$ and $E_C/E_J \ll 1$.

\begin{figure}[ht]
\rotatebox{0}{\resizebox *{\columnwidth}{6.8cm} {\includegraphics
{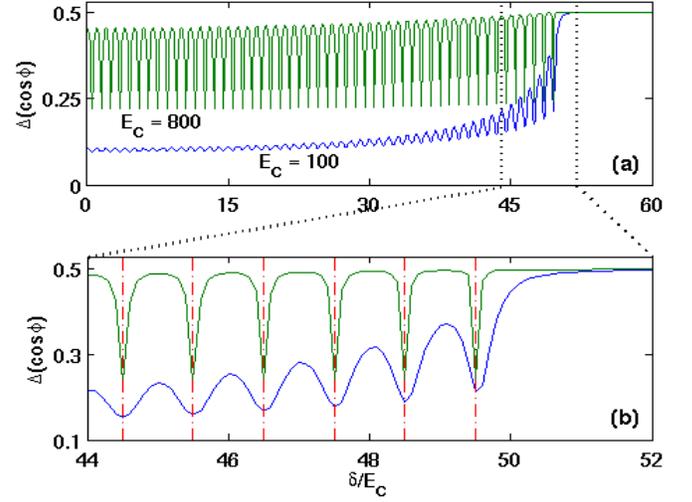}}} \caption{Coherence fluctuations
in the ground states of asymmetric systems with $N = 100$, $E_J =
100$ and large values of $E_C$. (a) Coherence fluctuations
$\Delta(\cos \phi)$ vs. $\delta/E_C$; and (b) the enlarged region
between two dot lines shown in (a). Blue and green lines
correspond to the values $E_C = 100$ and $E_C=800$, respectively.
Resonant suppression of the coherence fluctuations occurs at
$\delta = -E_C (k+1/2)$ for $k = (-N/2, -N/2 + 1, ..., N/2 -1)$,
marked by dashed red lines in (b).} \label{fig3}
\end{figure}

For asymmetric systems with large values of $E_C/E_J$ at zero
temperature, the coherence fluctuations can be suppressed
resonantly, due to degeneracy and quasi-degeneracy of the energy
levels occurring at some particular values of the ratio
$\delta/E_C$. For systems without tunneling, the eigenstate of
quasi-angular momentum $n$ has the energy $E^{*}(n) = \delta n + E_C
n^2/2$, and so that two neighboring levels become degenerate when
$E^{*}(n) = E^{*}(n+1)$. This means that, near the values of
$\delta/E_C = - (n+1/2)$ for $n = (-N/2, -N/2 + 1, \cdots, N/2 -1)$,
at least one atom can tunnel freely between two wells, even in the
limit of very weak tunneling (i.e., for very small values of $E_J$).
Therefore, the coherence fluctuations from inter-particle
interaction will be suppressed resonantly, with the resonance width
depending on $E_J$. This is a single-atom resonance effect caused by
the degeneracy and quasi-degeneracy between the states of the
neighboring angular momenta $n$. For very small values of $E_J/E_C$,
we perform a perturbation analysis by assuming that the perturbed
ground state is a superposition of the two lowest states of the
system without tunneling. Namely, the ground state $\left\vert
\psi\right\rangle_{G} = \alpha_{1}\left\vert L_{1}\right\rangle +
\alpha _{2}\left\vert L_{2}\right\rangle$, where $\left\vert
L_{1}\right\rangle$ and $\left\vert L_{2}\right\rangle$ are the two
lowest states. The two coefficients are determined by the lowest
energy state of the following eigenvalue equation,
$$
\left(
\begin{array}{cc}
H_{11} & H_{12} \\
H_{21} & H_{22}%
\end{array}%
\right) \left(
\begin{array}{c}
\alpha _{1} \\
\alpha _{2}%
\end{array}%
\right) = E \left(
\begin{array}{c}
\alpha _{1} \\
\alpha _{2}%
\end{array}%
\right) ,  \label{per}
$$
in which $H_{ij}=\left\langle L_{i}\right\vert H\left\vert
L_{j}\right\rangle $ ($i,j=1,2).$  The perturbation results confirm
the resonant suppression of coherence fluctuations, and they show
that the resonance width increases with $E_J/E_C$. Similarly, if we
prepare a state of a quasi-angular momentum $n$, the correlated
tunneling of an atomic pair (two-atom resonance) should correspond
to the degeneracy condition $E^{*}(n) = E^{*}(n+2)$. Our results for
single- and two-atom resonances explain consistently the recent
observations of conditional tunneling~\cite{ICOLS2007-Bloch}. In
Fig.~\ref{fig3}, we show the coherence fluctuations in the ground
states of asymmetric systems with $N = 100$, $E_J = 100$ and large
values of $E_C$. For a very large value of the charging energy, $E_C
= 800$ (green lines), our numerical results differ slightly from the
the results calculated by the perturbation analysis. Even for a
moderate value of the charging energy, $E_C = 100$, our numerical
results clearly show the resonant suppression of the coherence
fluctuations. Apparently, the resonance for the case $E_C = 100$ is
wider then that for the case $E_C = 800$.

\begin{figure}[ht]
\rotatebox{0}{\resizebox *{\columnwidth}{6.8cm} {\includegraphics
{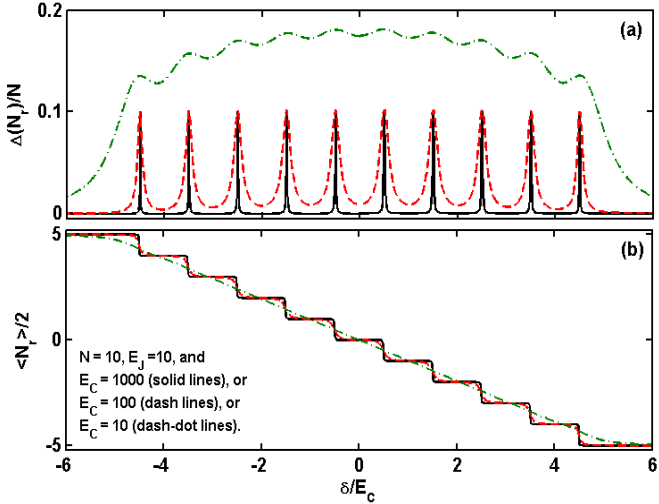}}} \caption{Peaks of single-atom
resonant tunneling and plateaus of interaction blockade. (a) The
fluctuations of relative particle numbers and (b) the expectation
values of half relative particle numbers versus $\delta/E_{C}$.
Solid, dash and dash-dot lines correspond to $E_C/E_J = 100$, $10$
and $1$, respectively.} \label{fig4}
\end{figure}

The single-atom resonant tunneling is a signature of interaction
blockade induced by the s-wave scattering between atoms. It is
well known that the Coulomb blockade of single-electron tunneling
occurs because of the charging energy \cite{Coulomb-blockade}. In
the Bose-Josephson junction (1), the asymmetry and the
inter-particle interaction act the roles of the applied voltage
and the charging interaction, respectively. As a results of the
interaction blockade, adiabatically sweeping the asymmetry, atoms
in the ground state tunnel through the potential barrier
one-by-one and the coherence vs. asymmetry dependence is a set of
sharp peaks between plateaus of interaction block. The interplay
of the single-atom resonant tunneling and the interaction blockade
allows one to create so called single-atom devices, which would
provide promising application in sensitive metrology and
information technology like their electronic counterparts
\cite{Coulomb-blockade-application,Coulomb-blockade-application-book}.
For a strongly interacting system in optical lattices with
asymmetry imposed by a harmonic potential, the appearance of
Mott-insulator shells \cite{MI-shells-MIT,MI-shells-Mainz} is also
the results of interaction blockade. In Fig.~\ref{fig4}, we show
resonant tunneling peaks and interaction blockade plateaus for a
quantized Bose-Josephson junction of $N = 10$, $E_J = 10$ and $E_C
= (1000, 100, 10)$. In which, $\Delta(N_r)$ and $<N_r>$ are the
quantum fluctuation and expectation of the relative particle
number $N_r$, respectively. The regular steps in $<N_r>$ and sharp
peaks in $\Delta(N_r)$ clearly show the single-atom resonant
tunneling and the interaction blockade. It also show that the
interaction blockade becomes more and more significant when the
interaction energy $E_C$ increases. In the recent experiment of
atoms confined in double-well lattices \cite{ICOLS2007-Bloch},
$N=2$ and $E_J / E_C \approx 0.2$. For the state of two atoms in
the first well, $n = - 1$, the single-atom (and two-atom)
resonances occur at $\delta = E_C/2 = 0.78 E_r$ (and $\delta = 0$)
when the resonant conditions $E^{*}(n) = E^{*}(n+1)$ (and
$E^{*}(n) = E^{*}(n+2)$) are satisfied. In which, the recoil
energy $E_r = h^2/(2m\lambda^2)$ with the mass $m$ for a
$^{87}\textrm{Rb}$ atom, the Planck's constant $h$, and the
short-lattice wave-length $\lambda = 765~\textrm{nm}$.

The analysis presented above describes the system properties at
zero temperature. For finite-temperature systems, assuming a
canonical ensemble statistics, the equilibrium states are mixed
states described by the density matrix $ \rho (T) = \exp(-\beta
H)/Tr[\exp(-\beta H)]$. Here, $\beta=1/(k_{B} T)$, $Tr[\rho
(T)]=1$, $k_{B}$ is the Boltzmann constant, and $H$ is the
Hamiltonian matrix. The thermodynamic average and variance of an
arbitrary operator $Q$ are $ \langle Q \rangle _{T} = Tr[\rho (T)
Q]$ and $\Delta (Q)_{T} = Tr[\rho (T) (Q - \langle Q
\rangle)^{2}]$, respectively.

\begin{figure}[ht]
\rotatebox{0}{\resizebox *{\columnwidth}{6.5cm} {\includegraphics
{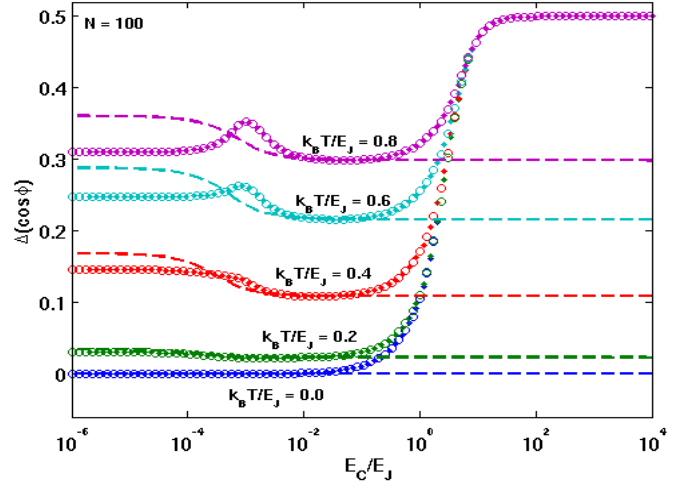}}} \caption{Quantum and thermal
fluctuations of the phase coherence vs. $E_C/E_J$ for symmetric
systems of $N=100$ and different values of temperature. Circles
and dots correspond to the cases $E_J = 1.0$ and $E_J=0.5$,
respectively. Dashed lines are estimated with the classical
Boltzmann theory.} \label{fig5}
\end{figure}

For the systems of fixed parameters, we find that the coherence
fluctuations grow with temperature. However, for the systems at
fixed temperature, the parametric dependence of the coherence
fluctuations relies on the values of temperature. For low
temperatures, the coherence fluctuations grow with $E_C/E_J$. This
finite-temperature behavior is qualitatively consistent with the
experimental data on noise thermometry with two Josephson-linked
condensates of Bose atoms~\cite{ThermalDWI-Heidelberg}. For higher
temperatures, the coherence fluctuations grow with $E_C/E_J$
except for the anomalous behaviors near the value $E_C/E_J \sim
4/N^2$. For every eigenstate, we numerically find that the quantum
fluctuations of coherence increase with $E_C/E_J$. This indicates
that the monotonous growth of the coherence fluctuations vs
$E_C/E_J$ (except for the anomalous region) in finite-temperature
systems is a signature of the quantum fluctuations.

Using the mean-field Hamiltonian (\ref{eq_H}) with the same
parameters, we estimate the coherence fluctuations with the
classical Boltzmann theory. For systems of low temperatures and
very small values of $E_C/E_J$, the classical estimations show
excellent consistence with the quantum ones. Otherwise, the
classical results greatly differ from the quantum ones. This means
that the mean-field Hamiltonian works good for systems of low
temperatures and very small values of $E_C/E_J$. Because the
mean-field treatment uses the coherent state approximation and
only the low-energy states for systems of small $E_C / E_J$ are
close to coherent states, the excellent consistence for systems of
low temperature and small values of $E_C / E_J$ is a natural
result of the mean-field approximation.

In Fig.~\ref{fig5}, we show the quantum and thermal fluctuations
of the phase coherence vs. $E_C/E_J$ for symmetric systems of
$N=100$ and different values of $T$. In which, the dashed lines
are obtained from the mean-field model (2) and the other data are
calculated from the full quantum model (1). The region near
$E_C/E_J \geq 4/N^2$ is the region of degenerated highest excited
states which corresponds to the macroscopic quantum self-trapping
in mean-field models~\cite{MQST-Smerzi,bifurcation-Lee,note}. The
anomalous behavior of the coherence fluctuations in this region is
caused by the occupation of the high excited states of two-fold
degeneracy when the temperature becomes sufficiently high. We also
calculate the averaged energy spacing for different parameters,
and the results show a minimum in the crossover region between
normal and self-trapping regions. For fixed and sufficiently high
temperatures, the populations of high-excited states in this
crossover region will larger than those in other regions, and then
the coherence fluctuations show an anomalous peak.

In conclusion, we have studied the mechanisms of coherence
fluctuations and interaction blockade in two Josephson-linked
Bose-Einstein condensates with repulsive interaction. Our results
for the coherence fluctuations and their resonant suppression are
qualitatively consistent with several recent experimental studies
of coherence
fluctuations~\cite{DWI-Austria,ThermalDWI-Heidelberg}, conditional
tunneling of single atoms and atom pairs~\cite{ICOLS2007-Bloch},
number squeezing~\cite{Science-squeezing-Kasevich}, and
squeezing-assisted long-time coherence~\cite{LTC-MIT}. We have
revealed how the fluctuations depend on the system parameters and
temperature. The different parametric dependencies of the
coherence fluctuations and the number squeezing provide a route
for controlling the coherence time experimentally by adjusting the
barrier height (junction energy)~\cite{LTC-MIT}. In particular, we
have shown that a competition between the potential asymmetry and
inter-particle interaction leads to the resonant tunneling and
interaction blockade. The resonant effects explain consistently
the recent observations of the conditional tunneling of single
atoms and atom pairs~\cite{ICOLS2007-Bloch}. More promisingly, the
mechanisms of resonant single-atom tunneling and interaction
blockade provide an alternative approach for designing single-atom
devices for applications in sensitive metrology and information
technology. At higher temperatures, an anomalous behavior of the
coherence fluctuations is an indicative of the energy level
degeneracy corresponding to the macroscopic quantum self-trapping
effect~\cite{MQST-Smerzi,bifurcation-Lee}.

\acknowledgments We thank Dr. Elena Ostrovskaya for stimulating
discussions. This work has been supported by the Australian
Research Council through the Discovery and Centre of Excellence
projects. Dr. L.-B. Fu is partially supported the National Natural
Science Foundation of China (NNSFC, Grants No. 10604009).


\begin{thebibliography}{0}
\bibitem{note}
Writing the pure states for the quantum system (1) as
$|\Psi\rangle = \sum _{n=-N/2}^{N/2} C(n) |n = N_r / 2 \rangle$,
the density distributions $|C(n)|^2$ vs. $E_C/E_J$ for the
highest-excited and sub-highest-excited states can be calculated.
Indeed, the classical $\pi$-phase stationary states in the
mean-field Hamiltonian (2) correspond to the extrema (minima or
maxima) of $|C(n)|^2$. In which, the stable and unstable fixed
points correspond to the maxima and minima, respectively. The
details of the correspondence between quantum degeneracy and
classical bifurcation will be published elsewhere.
\end{thebibliography}
\end{document}